\documentclass[12pt]{iopart}

\usepackage{graphicx}
\usepackage{amsfonts}

\usepackage{epsfig}
\usepackage{graphicx}
\usepackage{dcolumn}
\usepackage{bm}
\usepackage{times}
\usepackage{xcolor}

\begin{document}

\title{Simulated identification of epidemic threshold on finite-size networks}

\author{Panpan Shu$^{1}$, Wei Wang$^{1}$, Ming Tang$^{1,2}$, Younghae Do$^{3}$}
\address{$^{1}$ Web Sciences Center, University of Electronic
Science and Technology of China, Chengdu 610054, China}
\address{$^{2}$ State key Laboratory of Networking and Switching Technology, Beijing University of Posts and Telecommunications, Beijing 100876, China}
\address{$^{3}$ Department of Mathematics, Kyungpook National University, Daegu 702-701, South Korea}
\address{E-mail: tangminghuang521@hotmail.com}

\begin{abstract}
Epidemic threshold is one of the most important features of the epidemic dynamics. Through a lot of numerical simulations in classic Susceptible-Infected-Recovered (SIR) and Susceptible-Infected-Susceptible (SIS) models on various types of networks, we study the simulated identification of epidemic thresholds on finite-size networks. We confirm that the susceptibility measure goes awry for the SIR model due to the bimodal distribution of outbreak sizes near the critical point, while the simulated thresholds of the SIS and SIR models can be accurately determined by analyzing the peak of the epidemic variability. We further verify the accuracy of theoretical predictions derived by the heterogeneous mean-field theory (HMF) and the quenched mean-field theory (QMF), by comparing them with the simulated threshold of the SIR model obtained from the variability measure. The results show that the HMF prediction agrees very well with the simulated threshold, except the case that the networks are disassortive, in which the QMF prediction is more close to the simulated threshold.
\end{abstract}



 \maketitle
\tableofcontents
\section{Introduction} \label{sec:intro}

Models for disease propagation are the foundation of the study of spreading dynamics on complex networks~\cite{Barrat:2008,Albert:2012}. Two epidemic models of particular importance are the susceptible-infected-susceptible (SIS) and susceptible-infected-recovered (SIR) models~\cite{Anderson:1992}. At each time step, an infected node can transmit a disease to each of its susceptible neighbors with probability $\lambda$. At the same time, the infected nodes become susceptible again in the SIS model or recover in the SIR model with probability $\mu$. In the SIS model, a critical value of the effective transmission rate $\lambda/\mu$ separates the absorbing phase with only healthy nodes from the active phase with a stationary density of infected nodes. Differently, no steady state is allowed in the SIR model, but a threshold still exists above which the final fraction of infected nodes is finite~\cite{Pastor-Satorras:2014}.

The traditional theoretical study on the epidemic threshold of the SIS model was based on the heterogeneous mean-field (HMF) theory, which means that all the nodes within a given degree are considered to be statistically equivalent ~\cite{Dorogovtsev:2008RMP,Pastor-Satorras:2001PRE}. According to the HMF theory, the epidemic threshold of SIS model is given by~\cite{Pastor-Satorras:2001PRL,Boguna:2002PRE}
\begin{equation}\label{SIS_HMF}
    \lambda_c^{HMF}=\frac{\langle k \rangle}{\langle k^2 \rangle},
\end{equation}
where $\langle k \rangle$ and $\langle k^2 \rangle$ are the first and second moments of degree distribution $P(k)$~\cite{Newman:Networks}, respectively. On networks with power-law scaling $P(k)\sim k^{-\gamma}$~\cite{Newman:Networks,Albert:2002RMP}, where $\gamma$ is the degree exponent, one obtains a vanishing threshold in the thermodynamic limit for $\gamma \leq 3$, while the threshold is finite for $\gamma > 3$~\cite{Newman:2004CP}. As the quenched structure of the network and dynamical correlations between the state of
adjacent nodes are neglected in the HMF theory~\cite{Givan:2011JTB},
researchers proposed an important improvement over the HMF theory--- quenched mean-field (QMF) theory. The QMF theory fully preserves the actual quenched structure of the network described as its adjacency matrix, and the epidemic threshold is predicted to be ~\cite{Chakrabarti:2008ACM,Van Mieghem:2009ACM,Gomez:2010epl}
\begin{equation}\label{QMF}
    \lambda_c^{QMF}=\frac{1}{\Lambda_N},
\end{equation}
where $\Lambda_N$ is the maximum eigenvalue of the adjacency matrix of a given network. Given the scaling of $\Lambda_N$ with the maximum degree, $\Lambda_N\sim\{\sqrt{k_{max}}, \langle k^2\rangle/\langle k\rangle\}$~\cite{Chung:2003PNAS}, the epidemic threshold predicted by the HMF theory is the same as that from the QMF theory when $\gamma < 5/2$, while for $\gamma > 5/2$ the QMF prediction vanishes in the thermodynamic limit~\cite{Castellano:2010PRL}. Moreover, for a network with large size $N$, the more accurate SIS epidemic threshold
\begin{equation}\label{secondorder}
    \lambda_c^{(2)}=\lambda_c^{QMF}+o(\frac{\lambda_c^{QMF}}{N})
\end{equation}
is estimated by the second-order mean-field approximation~\cite{Cator:2012PRE}.

The earliest theoretical study on the SIR model is under
the assumption of homogeneous mixing, showing that the SIR epidemic threshold is inversely proportional to the average connectivity $\langle k \rangle$~\cite{Anderson:1992}.
At the HMF level~\cite{Barthelemy:2004PRL}, the epidemic threshold of SIR model takes the value
\begin{equation}\label{SIR_HMF}
    \lambda_c=\frac{\langle k \rangle}{\langle k^2 \rangle - \langle k \rangle}.
\end{equation}
The result of Eq.~(\ref{SIR_HMF}) coincides with the critical point of bond percolation, as the SIR model can be mapped to the bond percolation model~\cite{Newman:2002PRE}. According to the QMF theory, the epidemic threshold of SIR model has the same expression as Eq.~(\ref{QMF})~\cite{Chakrabarti:2008ACM}. For random networks without degree-degree correlations, Eq.~(\ref{QMF}) boils down to Eq.~(\ref{SIR_HMF})~\cite{Li:2012PRE}.

As the existing theories have inherent defects (e.g., the HMF theory neglects the quenched structure of the network, dynamical correlations are ignored in QMF theory)~\cite{Gleeson:2011PRL}, some numerical methods have been proposed to check the accuracy of the different theoretical estimations. Three conventional methods are finite-size scaling analysis~\cite{Marro:1999}, susceptibility~\cite{Binder: MCS}, and lifetime~\cite{Boguna:2013PRL}. Generally, the finite-size scaling analysis allows the precise numerical determination of the critical point in absorbing-state phase transitions (e.g., contact process and Ising model), but it can not estimate the transition point accurately for networks with strong structural heterogeneity~\cite{Ferreira:2011PRE,Hong:2007PRL}. So far the susceptibility method and lifetime method are only applied to the SIS model~\cite{Boguna:2013PRL,Ferreira:2012PRE}. Different from the case of the SIS model, the outbreaks change from an infinitesimal fraction ($\lambda<\lambda_c$) to a finite fraction ($\lambda\geq\lambda_c$) in the SIR model~\cite{Castellano:2012Srp}.
The widely accepted method for estimating the SIR epidemic threshold should be the percolation theory~\cite{Newman:2002PRE}, according to which the outbreak size is finite above the critical point. However, the critical value of the finite outbreak size can not be measured quantitatively in numerical simulations. Although the HMF theory has been indicated to be more accurate for predicting the epidemic threshold of SIR model in configuration model~\cite{Castellano:2010PRL}, the systematic investigation of the accurate determination of the SIR epidemic threshold is still lacking.

In this work, we perform a lot of numerical simulations of the SIR model on networks with finite size, and present a simulated method by analyzing the peak of the epidemic variability~\cite{crepey:2006PRE,Shu:2012} to determine the epidemic threshold. The accuracy of this method is checked by applying it on random regular networks (RRN), where the HMF is exact. The method is also employed to study the cases of scale-free networks and real networks.

We organize this paper as follows. In Sec.~\ref{sec:Numerical}, we describe the epidemic dynamics and present simulated method for determining epidemic threshold. In Sec.~\ref{sec:analysis}, we investigate some critical properties of the SIS and SIR dynamics, and discuss the validity of the simulated methods. The simulated thresholds of the SIR model on scale-free (SF) networks and real networks are discussed in Sec.~\ref{sec:applica}. Sec.~\ref{sec:conclusion} gives conclusions.

\section{An effective simulated identification measure} \label{sec:Numerical}

In simulations, we consider the SIS and SIR models for epidemics in discrete time. At the beginning, half of nodes are randomly
chosen as seeds in the SIS model. As the number of initial infected nodes affects the final outbreak size, we assume that only one node
is infected at the initial time in the SIR model. The simulations are implemented by using synchronous updating scheme. At each time step, each susceptible node $i$ becomes infected with probability $1-(1-\lambda)^{n_i}$ if it contacts with one or more infected neighbors, where $n_i$ is the number
of its infected neighbors. At the same time, all infected nodes are cured and become again susceptible at rate $\mu$ in the SIS model,
while they recover (or die) at rate $\mu$ and the recovered nodes acquire permanent immunity in the SIR model. Time is incremented by $\Delta t=1$,
and the SIS or SIR process is iterated with synchronous updating~\cite{Vespignani:2001PRE,Moreno:2011EPJB}. The SIS process ends after a long time step, and the SIR process ends when
there are no more infected nodes. Without lack of generality, we set $\mu=1$.

\begin{figure*}
\begin{center}
\epsfig{file=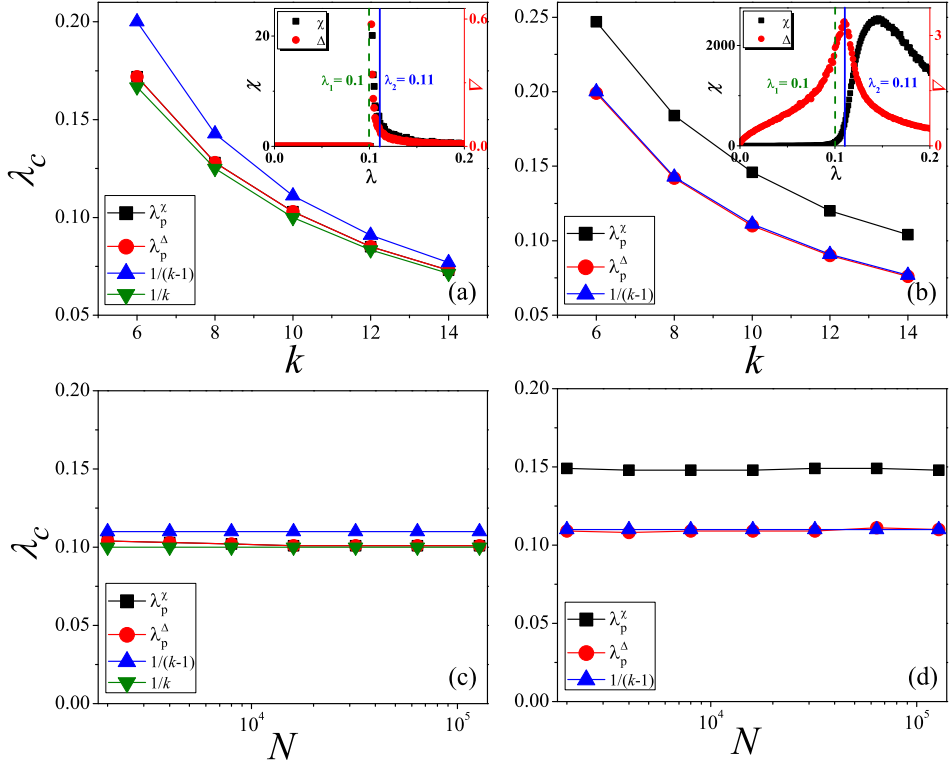,width=0.8\linewidth}
\caption{(Color online) Comparison of theoretical thresholds with simulated thresholds on RRNs.  The threshold $\lambda_c$ vs. degree $k$
for SIS (a) and SIR (b), where $N$ is set to $10^4$. The threshold $\lambda_c$ vs. network size $N$ for SIS (c) and SIR (d), where $k$ is set to $10$.
In each subfigure, ``squares", ``circles", ``triangleups" and ``triangledowns" denote $\lambda_p^{\chi}$,~$\lambda_p^{\Delta}$,~$1/(k-1)$ and $1/k$, respectively. Insets: Susceptibility $\chi$ and variability $\Delta$ as a function of $\lambda$. The results are averaged over $10^{4}$ independent realizations on a network.}

\label{fig:effctiveness}
\end{center}
\end{figure*}

For a RRN with constant degree $k$, the HMF predictions for the SIS and SIR models are accurate, namely $\lambda_c^{SIS}=1/k$ and $\lambda_c^{SIR}=1/(k-1)$~\cite{Dorogovtsev:2008RMP}, respectively. By comparing with the HMF predictions on RRNs, Figs.~\ref{fig:effctiveness} (a) and (b) check the accuracy of simulated threshold $\lambda_p^{\chi}$ from the \emph{susceptibility measure}
\begin{equation}\label{susceptibility}
\chi=N\frac{\langle \rho^2 \rangle - \langle \rho \rangle^2}{\langle \rho \rangle},
\end{equation}
where $\rho$ denotes the prevalence $\rho_I$ (i.e., the steady density
of infected nodes in the SIS model) or the outbreak size $\rho_R$ (i.e., the final density of recovered nodes in the SIR model). We find the SIS epidemic threshold determined by the susceptibility $\chi$ is very close to $\lambda_c^{SIS}=1/k$, but the simulated threshold of the SIR model is larger than $\lambda_c^{SIR}=1/(k-1)$. In other words, the susceptibility $\chi$ becomes invalid for estimating the epidemic threshold of the SIR model.

Here we employ the \emph{variability measure} $\Delta$~\cite{crepey:2006PRE,Shu:2012} to numerically determine the epidemic threshold:
\begin{equation}\label{variability}
\Delta=\frac{\sqrt{\langle \rho^2 \rangle - \langle \rho \rangle^2}}{\langle \rho \rangle},
\end{equation}
which can be explained as the standard deviation of the epidemic prevalence (or the outbreak size), and is a standard measure to determine critical point in equilibrium phase on magnetic system~\cite{Ferreira:2011PRE}. The insets of Figs.~\ref{fig:effctiveness} (a) and (b) show that the variability $\Delta$ reaches a maximum value, so we estimate the epidemic threshold from the position of the peak of the variability $\lambda_p^{\Delta}$. For the SIS model, we compare $\lambda_p^{\Delta}$ with the prediction from the HMF theory (i.e., $1/k$) and that from the pairwise approximation method (PA) (i.e., $1/(k-1)$)~\cite{Ferror:2014NJP} respectively [see Fig.~\ref{fig:effctiveness} (a)]. We find that the simulated threshold $\lambda_p^{\Delta}$ is consistent with the HMF prediction, which is almost the same as the $\lambda_p^{\chi}$. But for small $k$ it is smaller than the PA prediction which is more suitable for the SIS dynamics simulated by asynchronous updating~\cite{Ferreira:2012PRE}. With the increase of $k$, the gap between $\lambda_p^{\Delta}$ and PA prediction will decrease as $1/k \simeq 1/(k-1)$ for large $k$. Note that our synchronous updating scheme accounts for the difference between $\lambda_c^{SIS}=1/k$ in this work and $\lambda_c^{SIS}=1/(k-1)$ in Ref.~\cite{Ferreira:2012PRE}. For the SIR model, $\lambda_p^{\Delta}$ is always consistent with the HMF prediction $\lambda_c^{SIR}=1/(k-1)$. To make a further comparison with the susceptibility measure, we consider the relationship between the epidemic threshold and network size in Figs.~\ref{fig:effctiveness} (b) and (d). Once the degree $k$ is given, the simulated thresholds $\lambda_p^{\chi}$ and $\lambda_p^{\Delta}$ do not change with network size $N$, and $\lambda_p^{\Delta}$ is closer to $\lambda_c^{SIR}=1/(k-1)$. From the above, we know that the variability $\Delta$ performs well in both the SIS model and the SIR model, while the susceptibility $\chi$ only can work in the SIS model. Thus, a new problem has arisen: why the variability $\Delta$ performs well but the susceptibility $\chi$ goes awry for the SIR model?

\section{Analysis of simulated identification measure near the critical point} \label{sec:analysis}

\subsection{Comparison of epidemic outbreak distribution in the SIS and SIR models}\label{subsec:distribution}

\begin{figure}
\begin{center}
\epsfig{file=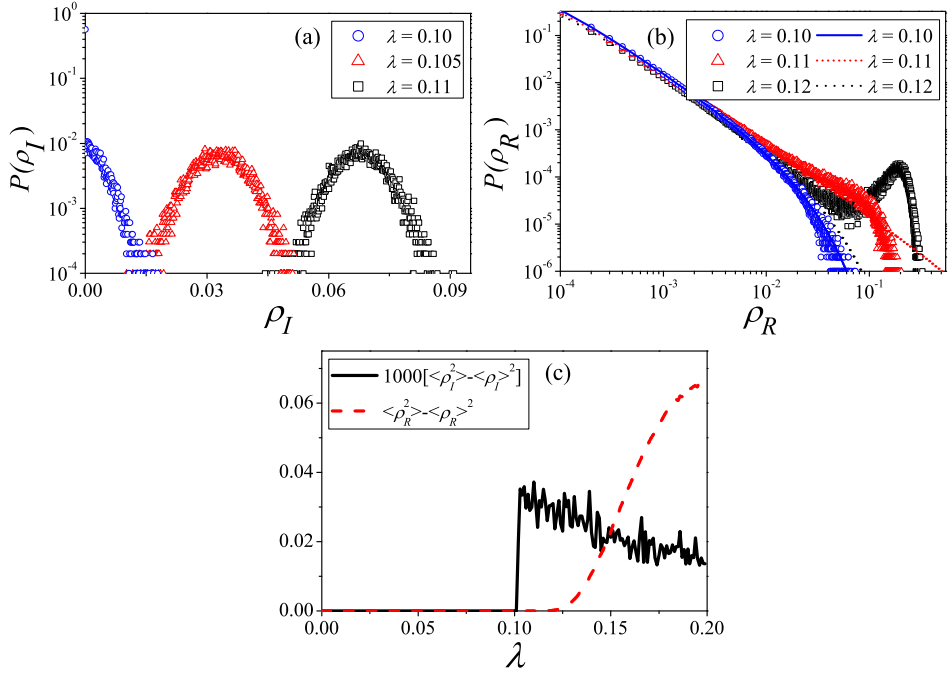,width=0.8\linewidth}
\caption{(Color online) Critical distribution and fluctuations of epidemic outbreaks on a RRN.
(a) Simulated distribution of the prevalence $\rho_I$ in SIS model for $\lambda=0.10$
(circles), $\lambda=0.105$ (triangles), and $\lambda=0.11$ (squares).
(b) Simulated distribution of outbreak sizes $\rho_R$ in SIR model for $\lambda=0.10$ (circles),
$\lambda=0.11$ (triangles), and $\lambda=0.12$ (squares), where blue solid, red short
dash and black dot lines respectively represent the theoretical distributions given by Eq.~(\ref{eq:p4}). (c) Fluctuations of the
prevalence $1000(\langle \rho_I^2 \rangle - \langle \rho_I \rangle^2)$ (solid line) and the outbreak size
$\langle \rho_{R}^{2} \rangle - \langle \rho_{R} \rangle^2$ (dot line). The paraments are chosen as $N=10^4$
and $k=10$. The results are averaged over $10^{6}$ independent realizations on a network.
}
\label{fig:pr}
\end{center}
\end{figure}

To deal with that problem illuminated in Sec.~\ref{sec:Numerical}, we investigate the distribution of the epidemic prevalence
$\rho_I$ (the outbreak size $\rho_R$) and its fluctuation $\zeta=\langle \rho^2 \rangle - \langle \rho \rangle^2$ in the SIS (SIR) model. Fig.~\ref{fig:pr} shows these results on
a RRN with $k=10$. We see that the distribution of the prevalence near the SIS epidemic threshold is very different from
the outbreak size distribution near the epidemic threshold of SIR model.

For the SIS model in Fig.~\ref{fig:pr} (a), we obtain the simulated threshold $\lambda_c =1/\langle k \rangle\simeq 0.1$. Below the threshold (i.e., $\lambda<\lambda_c$), a nonzero $\rho_I$ can hardly exist, since the disease will eventually die out. At the threshold (i.e., $\lambda=0.1$), although the prevalence is close to be an exponential distribution, the probability of $\rho_I=0$ is maximum,
which means the prevalence is still very small. Above the threshold (e.g., $\lambda$=0.105 and 0.11), the prevalence approximates a normal distribution, where the position of the peak value is determined by the average density of infected nodes $\langle\rho_I\rangle$. Fig.~\ref{fig:pr} (c) shows that the fluctuation of $\rho_I$ in SIS model is on the order of one-thousandth of the $\rho_R$ fluctuation in SIR model. When $\lambda < \lambda_c$, $\zeta$ is zero, and the corresponding susceptibility $\chi$ and variability $\Delta$ are zero. When $\lambda \geq \lambda_c$, $\zeta$ abruptly becomes a finite value and changes little with $\lambda$, while $\langle\rho_I\rangle$ increases with $\lambda$. As a result, the peaks of the susceptibility $\chi$ and the variability $\Delta$ appear at the same $\lambda\simeq\lambda_c$ [see the inset of Fig.~\ref{fig:effctiveness} (a)], which is consistent with the HMF prediction.

For the SIR model, the variability $\Delta$ determines the simulated threshold $\lambda_c = 1/(\langle k \rangle-1)\simeq0.11$. In Fig.~\ref{fig:pr} (b), the outbreak sizes follow approximately an exponential distribution at $\lambda=0.1$. Near the critical point $\lambda\simeq\lambda_c$, the outbreak sizes follow a power-law distribution $P(\rho_R)\sim \rho_R^\alpha$ with a cutoff at some value, where $\alpha\simeq-1.5$~\cite{Ben-Naim:2004PRE,Ben-Naim:2012EPJB,Kessler:2007PRE}. Since the disease may die out quickly or infect a subset of nodes when $\lambda>\lambda_c$, the distribution of outbreak sizes is bimodal~\cite{Zanette:2001PRE,Khalleque:2013JPA}, with two peaks occurring at $\rho_R=1/N$ and $\rho_R\simeq0.2$ at $\lambda=0.12$, respectively. Therefore, the fluctuation of the outbreak sizes increases monotonically with $\lambda$ above the critical point in Fig.~\ref{fig:pr} (c).

Moreover, the theoretical distribution of the small epidemic sizes (see Appendix) is in good agreement with the results obtained by numerical simulations in Fig.~\ref{fig:pr}~(b). The theoretical probability from Eq.~(\ref{eq:p4}) is consistent with the simulated results for relatively small outbreak size ($\rho_R<0.05$). Near the critical point, the theoretical results prove that the outbreak sizes indeed obey a power-law distribution with the exponent -1.5. When $\lambda>\lambda_c$, some large outbreak sizes constitute a lump in the simulated scattergram, but the probability of large outbreak sizes can not be solved from Eq.~(\ref{eq:p4}). We thus speculate
that the non-ignorable lump may be influential in simulated determination of SIR epidemic threshold.

\subsection{Effectiveness of simulated identification measure under cutoff hypothesis}\label{subsec:hypothesis}
\begin{figure}[b]
\begin{center}
\epsfig{file=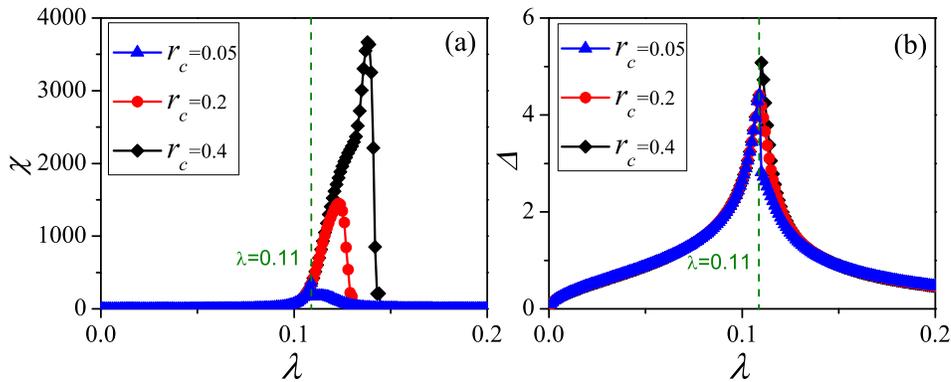,width=0.8\linewidth}
\caption{(Color online) Susceptibility $\chi$ and variability $\Delta$ with cutoff as a function of
$\lambda$ on a RRN. (a) $\chi$ vs. $\lambda$, where only the small
outbreak sizes with $\rho_R\leq r_c$ are considered when
$\lambda > \lambda_c$. (b) $\Delta$ vs.~$\lambda$, where the theoretical distribution of the lump is assumed
to be a Dirac delta function. ``triangles", ``circles" and ``diamonds" denote cutoff values $r_c$ = 0.05, 0.2 and 0.4,
respectively. The paraments are chosen as $N=10^4$ and $k=10$. The results are averaged over $10^{6}$ independent realizations on a network.
}
\label{fig:analysis}
\end{center}
\end{figure}

To verify the rationality of the speculation, Fig.~\ref{fig:analysis} investigates the effectiveness of the
variability and susceptibility measures under some cutoff hypothesis. We set the cutoff value of the outbreak
size as $r_c$, which means the outbreak sizes larger than $r_c$ are excluded in Fig.~\ref{fig:pr} (b). Three kinds of $r_c$ are considered,
where $r_c=0.05$ corresponds to the maximum value of small outbreak size before the lump appears in the simulated distribution, $r_c=0.2$
means that the distribution consists of a part of the lump, and $r_c=0.4$ means that there is a complete lump in the
distribution. When calculating the susceptibility in Fig.~\ref{fig:analysis} (a), all possible outbreak sizes are considered for $\lambda \leq \lambda_c$, while only the outbreak size with $\rho_R \leq r_c$ is required at $\lambda > \lambda_c$. The susceptibility measure can indeed give a quite accurate estimate of the SIR epidemic threshold when the whole lump is ignored (i.e., $r_c=0.05$). With the increase of $r_c$, the peak position of the susceptibility $\chi$ gradually shifts to the right for large outbreak sizes are considered. This indicates that the susceptibility $\chi$ lose its effectiveness on determining the SIR epidemic threshold due to the existence of the lump.

We have found from simulations that the cutoff value $r_c$ does not affect the simulated threshold $\lambda_p^{\Delta}$ corresponding to the first peak of $\Delta$. Then, the effectiveness of the variability $\Delta$ is further checked in theory.
As the simulated distribution of the large outbreak
sizes is concentrated, we assume the probability distribution of the lump is a Dirac delta function in theory.
That is to say, there is a lump located at $r=r_c$ with $P(r_c)=1-\Sigma_{\rho_R<r_c} P(\rho_R)$ in the theoretical
probability distribution diagram of outbreak sizes. Then, we plot the variability measure as a function of
$\lambda$ for different values of $r_c$ in Fig.~\ref{fig:analysis}~(b). The variability $\Delta$ measures
the heterogeneity of the outbreak sizes distribution, which is strongest at the critical point~\cite{Ben-Naim:2004PRE,Ben-Naim:2012EPJB,Kessler:2007PRE}. Therefore, the peak position
of the variability measure does not change with the size of the lump, as shown in Fig~\ref{fig:analysis}(b).

From the above analysis, we can conclude that the variability $\Delta$ is effective in determining the
epidemic threshold of SIR model, while the bimodal distribution of outbreak sizes for $\lambda>\lambda_c$ leads to the
obvious difference between the HMF prediction and the simulated threshold from the susceptibility $\chi$.

\section{Applications of simulated identification method}
\label{sec:applica}

In this section, we discuss the accuracy of the theoretical estimations from the HMF theory and from the QMF theory on both scale-free and real networks, by comparing them with the simulated threshold from the variability $\Delta$.

\subsection{Comparison of SIR epidemic thresholds on scale-free networks} \label{subsec:SF}

\begin{figure*}[b]
\begin{center}
\epsfig{file=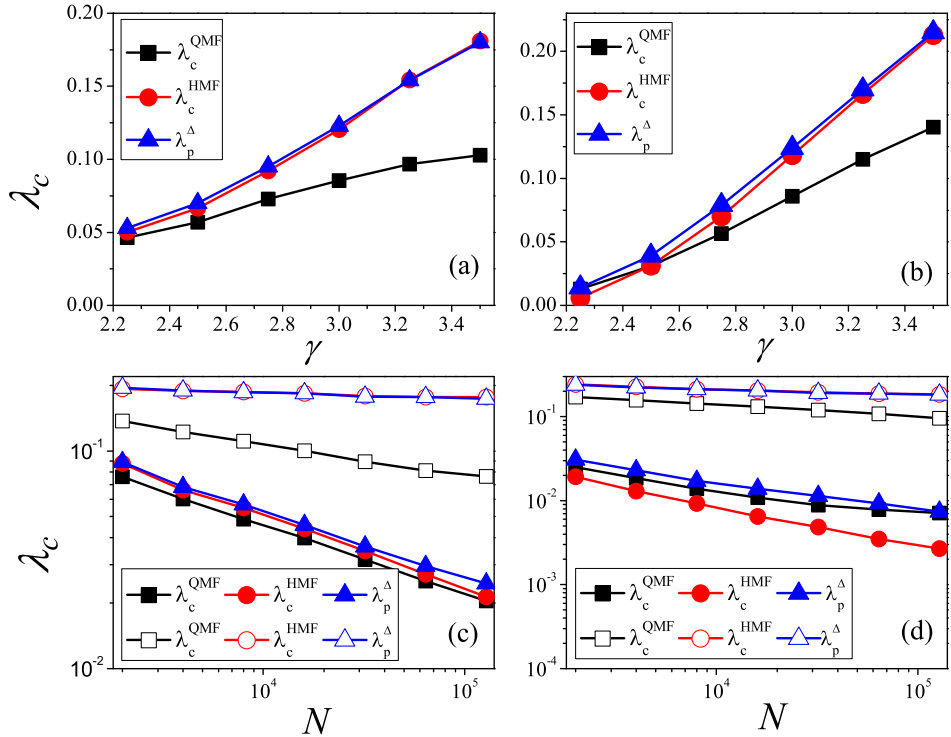,width=0.8\linewidth}
\caption{(Color online) Comparison of theoretical thresholds with simulated thresholds on SFNs. $\lambda_c$ vs.
$\gamma$ on SFNs with structural cutoff (a) and natural cutoff (b), where $N$ is set to $10^4$. $\lambda_c$ vs.
$N$ on SFNs with structural cutoff (c) and natural cutoff (d), where solid and empty symbols denote $\gamma=2.25$
and 3.50, respectively. ``squares", ``circles" and ``triangles" denote $\lambda_c^{QMF}$, $\lambda_c^{HMF}$
and $\lambda_p^{\Delta}$, respectively. The results are averaged over $10\times10^{4}$ independent realizations on different $10$ networks.}
\label{fig:SFN}
\end{center}
\end{figure*}

We first build scale-free networks (SFNs) with degree distribution $P(k)\sim k^{-\gamma}$ based on the
configuration model~\cite{Newman:Networks}. The so-called structural cutoff $k_{max}\sim N^{1/2}$ and natural
cutoff $k_{max}\sim N^{1/{\gamma-1}}$~\cite{Boguna:2004EPJB} are considered to constrain the maximum possible
degree $k_{max}$ on SFNs. We consider the SIR model on SFNs with structural cutoff in
Figs.~\ref{fig:SFN} (a) and (c), where the SIR epidemic threshold increases monotonically with the degree exponent
$\gamma$ and decreases linearly with the network size $N$~\cite{Binder: MCS}. When the structural cutoff
makes the degree-degree correlations vanish~\cite{Boguna:2004EPJB}, the HMF prediction $\lambda_c^{HMF}$
is much close to the simulated threshold $\lambda_p^\Delta$, while there is an obvious difference between the
QMF prediction $\lambda_c^{QMF}$ and $\lambda_p^\Delta$. According to Ref.~\cite{Lee:2013PRE},
the epidemic threshold is related to the largest degree $k_{max}$, whose variation with $N$ depends strongly
on $\gamma$. Thus, $\lambda_c$ drops rapidly for $\gamma=2.25$ and changes slowly with $N$ for $\gamma=3.5$
[see Fig.~\ref{fig:SFN} (c)].

The SFNs with natural cutoff are considered in Figs.~\ref{fig:SFN}
(b) and (d), where the variations of epidemic threshold with $\gamma$ and $N$ are similar to the result on SFNs
with structural cutoff. The HMF prediction performs an accurate prediction but there is a gap between the QMF prediction and the
simulated threshold when $\gamma>3$. Since the disassortative degree-degree correlations exist when $\gamma<3$,
there is a slight difference between $\lambda_c^{HMF}$ and $\lambda_p^\Delta$. Specially, Fig.~\ref{fig:SFN}
(d) shows a more clear distinction between $\lambda_c^{HMF}$ and $\lambda_p^\Delta$ for SFNs with natural
cutoff when $\gamma=2.25$, while the QMF prediction is very close to the simulated threshold for the
principle eigenvector is delocalized when $2<\gamma\leq5/2$~\cite{Goltsev:2012PRL}. It can be seen from the above analysis,
the prediction of the HMF theory seems to be much more accurate than the QMF prediction in most cases on SFNs~\cite{Castellano:2010PRL}.

\subsection{Comparison of epidemic thresholds on real networks} \label{subsec:real}

\begin{figure*}[h]
\begin{center}
\epsfig{file=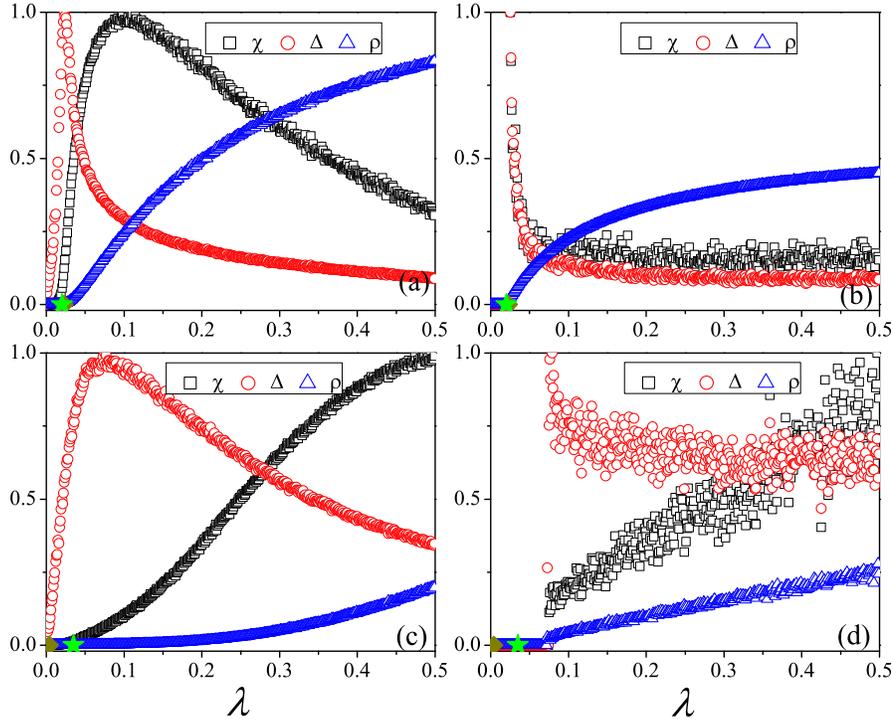,width=0.75\linewidth}
\caption{(Color online) Susceptibility $\chi$ and variability $\Delta$ as a function of $\lambda$ on real networks.
$\chi$, $\Delta$ and $\rho$ vs. $\lambda$ for SIR (a) and SIS (b) on Hamsterster full network.
$\chi$, $\Delta$ and $\rho$ vs. $\lambda$ for SIR (c) and SIS (d) on Facebook (NIPS) network. ``squares", ``circles"
and ``triangles" denote $\chi$, $\Delta$ and $\rho$, respectively. ``green star'' denotes
$\lambda_c^{QMF}=1/\Lambda_N$, ``yellow diamond'' denotes $\lambda_c^{HMF}=\langle k \rangle/[\langle k^2 \rangle-\langle k \rangle]$ in (a) and
(c), and $\lambda_c^{HMF}=\langle k \rangle/\langle k^2 \rangle$ in (b) and (d).
The susceptibility $\chi$ and variability $\Delta$ are normalized with $\chi_{max}$ and $\Delta_{max}$, respectively. The results are averaged over $10^{4}$ independent realizations on each network.}

\label{fig:real}
\end{center}
\end{figure*}

To further check the performances of the susceptibility $\chi$ and variability $\Delta$, Fig.~\ref{fig:real}
depicts $\chi$ and $\Delta$ as a function of $\lambda$ on Hamsterster full (containing friendships and family links between users of
the website hamsterster.com) and Facebook (NIPS) (containing Facebook user-user friendships) networks.
The simulated results intuitively show that the variability $\Delta$ always reaches a maximum value near the critical
point of $\rho$ (i.e., $\lambda_c$) for both SIS and SIR models. However, the peak of the susceptibility $\chi$
appears at a larger $\lambda$ in the SIR model, which is similar to the results in Sec.~\ref{sec:Numerical}.
The theoretical predictions of the HMF theory and of the QMF theory are quite close to the simulated threshold
determined by $\Delta$ on Hamsterster full network, which is assortative, but they become poor on Facebook (NIPS)
network, which is disassortative.

More detailed comparisons between the simulated and theoretical thresholds on real networks are presented in
Table~\ref{table}. For the SIR model, the simulated thresholds determined by the susceptibility [i.e.,
$\lambda_p^\chi(SIR)$] are greater than that obtained by the variability measure [i.e., $\lambda_p^\Delta(SIR)$].
Although the HMF prediction and the simulated threshold $\lambda_p^\Delta(SIR)$ are nearly the same for
assortative networks, there is an obvious difference between them for the networks showing significant
disassortative mixing. The QMF prediction is relatively worse than the HMF prediction for assortative networks,
but the former is close to $\lambda_p^\Delta(SIR)$ for some disassortative networks (e.g., Router views, CAIDI,
and email contacts). The two simulated thresholds of the SIS model, i.e., $\lambda_p^\chi(SIS)$ and
$\lambda_p^\Delta(SIS)$, are nearly the same for most of the real networks. For most of the assortative networks, the HMF prediction for the SIS model is very close to the simulated threshold. By calculating the inverse participation ratio IPR$(\Lambda)$
of real networks~\cite{Goltsev:2012PRL}, we see that, the QMF prediction agrees well with the simulated thresholds
of the SIS model when IPR$(\Lambda)\rightarrow0$ [i.e., the principal eigenvector of the adjacency matrix of a network $f(\Lambda)$ is
delocalized], but becomes poor when IPR$ (\Lambda)$ is large [i.e., the eigenvector $f(\Lambda)$ is localized].
This result agrees with the conclusion of Ref.~\cite{Goltsev:2012PRL} to a certain extent.

\begin{table*}[h]
\caption{Topology characteristics and epidemic thresholds of real networks. $N$ is the network size, $k_{max}$ is the maximum degree, $r$ is the degree correlations, $\lambda_c^{HMF}$(SIS) is the HMF result for SIS model, $\lambda_c^{HMF}$(SIR) is the HMF result for SIR model, and $\Lambda_N$ is the largest eigenvalue of adjacent matrix.
}\label{table}
  \resizebox{\textwidth}{!}{
  \begin{tabular}{|l|c|c|c|c|c|c|c|c|c|c|c|}
    \hline
    \hline
    Network & $N$ & $k_{max}$ & $r$ & $\lambda_c^{HMF}$(SIS) & $\lambda_c^{HMF}$(SIR) & $\lambda_c^{QMF}$ & $\lambda_p^\Delta$(SIR) & $\lambda_p^\chi$(SIR) & $\lambda_p^\Delta$(SIS) & $\lambda_p^\chi$(SIS) & IPR$(\Lambda_N)$ \\
    \hline
    Hamsterster full~\cite{Konect data} & 2000 & 273 & 0.023 & 0.023 & 0.023 & 0.020 & 0.023 & 0.108 & 0.025 & 0.025 & 0.009 \\
    Brightkite~\cite{Brightkie} & 56739 & 1134 & 0.010 & 0.016 & 0.016 & 0.010 & 0.014 & 0.238 & 0.012 & 0.012 & 0.006 \\
    arXiv astro-ph~\cite{astroph} & 17903 & 504 & 0.201 & 0.015 & 0.015 & 0.011 & 0.012 & 0.09 & 0.012 & 0.012 & 0.004 \\
    Pretty Good Privacy~\cite{Boguna:PRE2004} & 10680 & 206 & 0.239 & 0.053 & 0.056 & 0.024 & 0.053 & 0.477 & 0.033 & 0.033 & 0.017 \\
    US power grid~\cite{Watts:Nature1998} & 4941 & 19 & 0.003 & 0.258 & 0.348 & 0.134 & 0.446 & 0.496 & 0.261 & 0.264 & 0.041 \\
    Euroroad~\cite{Subelj:EPJB2011} & 1039 & 10 & 0.090 & 0.324 & 0.479 & 0.249 & 0.498 & 0.711 & 0.331 & 0.331 & 0.049 \\
    Facebook(NIPS)~\cite{Konect data} & 2888 & 769 & -0.668 & 0.004 & 0.004 & 0.036 & 0.075 & 0.494 & 0.079 & 0.497 & 0.244 \\
    Route views~\cite{Router} & 6474 & 1458 & -0.182 & 0.006 & 0.006 & 0.022 & 0.037 & 0.345 & 0.034 & 0.496 & 0.087 \\
    CAIDA~\cite{Router} & 26475 & 2628 & -0.195 & 0.004 & 0.004 & 0.014 & 0.019 & 0.336 & 0.019 & 0.019 & 0.024 \\
    email contacts~\cite{Kitsak:2010NatPhys} & 12625 & 576 & -0.387 & 0.009 & 0.009 & 0.02 & 0.027 & 0.404 & 0.024 & 0.025 & 0.013\\
    \hline
    \hline
  \end{tabular}   }
\end{table*}

\section{Conclusions} \label{sec:conclusion}

In summary, we have studied the simulated identification of epidemic threshold on complex networks with finite
size. First, the accuracies of the susceptibility and variability measures are checked by applying them on RRNs,
in which the HMF is exact. We have shown that the variability $\Delta$ is valid for determining the
simulated thresholds of the SIS and SIR models, while the susceptibility $\chi$ gives a larger SIR epidemic threshold.

In order to get a deep understanding of the two estimation methods, we have analyzed the epidemic spreading near
the critical point $\lambda_c$. For the SIS model, the epidemic quickly dies out when $\lambda<\lambda_c$. When
$\lambda\simeq\lambda_c$, although the prevalence approximates an exponential distribution, the probability of
$\rho=0$ is still maximum. Above the threshold with $\lambda > \lambda_c$, the prevalence is distributed homogeneously.
For the SIR model, the outbreak sizes follow approximately an exponential distribution when $\lambda < \lambda_c$.
At the critical point, the outbreak sizes follow a power-law distribution with the exponent -1.5. When
$\lambda\rightarrow\lambda_c^+$, the simulated distribution of outbreak sizes is bimodal with two peaks
occurring at $\rho=1/N$ and $O$(1). The probability of small outbreak sizes in theory is consistent with
that obtained by numerical simulations, but the probability of large outbreak sizes that constitute a lump
in the simulated scattergram can not be obtained theoretically. Based on a reasonable cutoff hypothesis,
we find the susceptibility measure can give a quite accurate SIR epidemic threshold when the second lump is ignored. Since the
variability measure reflects the relative fluctuation of epidemic spreading, it is always effective in determining
the epidemic threshold, where the distribution of outbreak sizes has a very strong heterogeneity.

Moreover, the simulated thresholds of the SIR model are investigated on scale-free and real networks.
All results indicate that the epidemic threshold determined by the variability $\Delta$ is more accurate than that
from the susceptibility $\chi$. The HMF prediction is in general more accurate, but it becomes worse due to the
existence of disassortative mixing on SFNs with natural cutoff and $\gamma<5/2$. Similarly, the HMF
approximation is accurate for the SIR model on real networks with assortative mixing, while it becomes very poor
for disassortive networks. We further confirm that although the QMF predictions is not accurate enough on assortative
it is valid for some disassortive networks.

We here put forward an estimation method, whose effectiveness has been verified by analyzing the critical
distribution. This method can be applied to the precise determination of epidemic threshold on various networks, and
could be extended to other dynamic processes such as information diffusion and behavior spreading. Further work should
be done to check the effectiveness of this method on more complicated networks (e.g., temporal networks~\cite{Holme:PR2012} and
multilayer networks~\cite{multiple ntwork}), and the cases in asynchronous updating scheme also need to be investigated. Besides, the
accurate analytic approximation of the epidemic threshold for general networks remains an important problem. This
work helps to verify theoretical analysis of critical point and would promote further study on phase transition of
epidemic dynamics.

\section*{Acknowledgements}
This work was partially supported by National Natural Science Foundation of China (Grant Nos. 11105025, 91324002),
China Postdoctoral Science Special Foundation (Grant No. 2012T50711), the Program of Outstanding Ph. D. Candidate in Academic Research by UESTC
(Grand No. YXBSZC20131033) and Open Foundation of State key Laboratory of Networking and Switching Technology (Beijing University of Posts and
Telecommunications) (SKLNST-2013-1-18). Y. Do was supported by Basic Science Research Program through the National Research Foundation of Korea (NRF) funded by the Ministry of Education, Science and Technology (NRF-2013R1A1A2010067).

\section*{Appendix}

For the case of the SIR model and similar models with no steady-state, the static properties (e.g., the final outbreak size and the critical point) of the epidemic outbreak can be mapped into a suitable bond percolation problem. In this framework, the distribution of occupied cluster sizes is related to the distribution of outbreak sizes. To get the distribution of small outbreak size in the SIR model with a fixed value of $\lambda$ when recovery rate $\mu=1$, we will present the derivation of the distribution of small occupied cluster sizes in bond percolation with bond occupation probability $\lambda$~\cite{Newman:2002PRE}.

After the percolation process on a general network with arbitrary degree distribution $p_k$, the average degree of the occupied network $A_1$, which composes of vertices and occupied edges, is $\langle k_T \rangle = \lambda \langle k \rangle$, where $\langle k \rangle$ is the average degree of the original network $A_0$. And the size distribution of the small subgraphs of network $A_1$ is
\begin{equation}\label{eq:p1}
    \pi_s = \frac{\langle k_T \rangle}{(s-1)!}[\frac{d^{s-2}}{d z^{s-2}}[g_1 (z)]^s]_{z=0},
\end{equation}
where s is the small subgraphs size and $g_1 (z)$ is the generating function of the excess degree of network $A_1$. In addition, the generating function of degree distribution of $A_1$ is
\begin{equation*}
    g_0(z)=\sum_{k=0}^{\infty}p_k(1-\lambda+z\lambda)^{k},
\end{equation*}
and we thus have
\begin{equation*}
    g_1(z)=\frac{g_0^{'}(z)}{g_0^{'}(1)}
\end{equation*}
In a random regular network, which has an unique degree $k$ with $p_k=1$, we can easily obtain that
\begin{equation}\label{eq:p2}
    g_0(z)=[1+(z-1)\lambda]^k,
\end{equation}
and
\begin{equation}\label{eq:p3}
    g_1(z)=[1+(z-1)\lambda]^{k-1}.
\end{equation}
Substituting Eq.~(\ref{eq:p3}) into Eq.~(\ref{eq:p1}), we can obtain the distribution of small outbreak
sizes of the disease as follow:
\begin{equation}\label{eq:p4}
    \pi_s = \frac{k \Gamma(a_2)}{\Gamma(a_0) \Gamma(a_1)}~\lambda^{s-1}(1-\lambda)^{s(k-1)-(s-2)},
\end{equation}
where $\Gamma(x+1)=x!, a_0=(s-2), a_1=s(k-1)-(s-1)$, and $a_2=s(k-1)-1$.

\end{document}